\documentclass[conference]{IEEEtran}
    \IEEEoverridecommandlockouts
    \pdfoutput=1
    \usepackage{cite}
    \usepackage{amsmath,amssymb,amsfonts}
    \usepackage{graphicx}
    \usepackage{textcomp}
    \usepackage{xcolor}
	\usepackage{booktabs}
    
    \usepackage{tikz}
    \usetikzlibrary{positioning,intersections,through,backgrounds,graphs,matrix,shapes,arrows}
    \usepackage{lmodern}
    \usepackage{booktabs} 
    \usepackage{multirow}
    \usepackage{listings}
    \usepackage{float}
    \newfloat{lstfloat}{htbp}{lop}
    \definecolor{darkgreen}{rgb}{0.1, 0.7, 0.3}
    \usepackage{enumitem}
    \usepackage{algorithm}
    \usepackage{todonotes}
    \usepackage[T1]{fontenc}
    \usepackage[noend]{algpseudocode}
    \usepackage{pdfpages}

    \setlist[itemize]{noitemsep, topsep=0pt}
    \definecolor{blue-main}{rgb}{0,0,1}
    \definecolor{dkgreen}{rgb}{0,0.6,0}
    \definecolor{gray}{rgb}{0.5,0.5,0.5}
    \definecolor{mauve}{rgb}{0.58,0,0.82}
    
    \def\BibTeX{{\rm B\kern-.05em{\sc i\kern-.025em b}\kern-.08em
    T\kern-.1667em\lower.7ex\hbox{E}\kern-.125emX}}
\begin{document}

\title{Asynchronous Execution of Python Code on Task-Based Runtime Systems \\
}

\author{\IEEEauthorblockN{R. Tohid\IEEEauthorrefmark{1},
Bibek Wagle\IEEEauthorrefmark{1},
Shahrzad Shirzad\IEEEauthorrefmark{1},
Patrick Diehl\IEEEauthorrefmark{1}, \\
Adrian Serio\IEEEauthorrefmark{1},
Alireza Kheirkhahan\IEEEauthorrefmark{1},
Parsa Amini\IEEEauthorrefmark{1}, \\
Katy Williams\IEEEauthorrefmark{2},
Kate Isaacs\IEEEauthorrefmark{2},
Kevin Huck\IEEEauthorrefmark{3},
Steven Brandt \IEEEauthorrefmark{1}  and
Hartmut Kaiser\IEEEauthorrefmark{1}}
\IEEEauthorrefmark{1} Louisiana State University,
\IEEEauthorrefmark{2} University of Arizona,
\IEEEauthorrefmark{3} University of Oregon \\
E-mail: $\{$mraste2, bwagle3, sshirz1, patrickdiehl, akheir1$\}$@lsu.edu, $\{$hkaiser, aserio, sbrandt, parsa$\}$@cct.lsu.edu,\\ khuck@cs.uoregon.edu, kisaacs@cs.arizona.edu, kawilliams@email.arizona.edu \\
URL: Patrick Diehl (https://orcid.org/0000-0003-3922-8419)
}

\maketitle

\begin{abstract}
    Despite advancements in the areas of parallel and distributed computing, the
    complexity of programming on High Performance Computing (HPC) resources has
    deterred many domain experts, especially in the areas of machine learning
    and artificial intelligence (AI), from utilizing performance benefits of
    such systems. Researchers and scientists favor high-productivity languages
    to avoid the inconvenience of programming in low-level languages and costs
    of acquiring the necessary skills required for programming at this level. In
    recent years, Python, with the support of linear algebra libraries like
    NumPy, has gained popularity despite facing limitations which prevent this
    code from distributed runs. Here we present a solution which maintains both
    high level programming abstractions as well as parallel and distributed
    efficiency. Phylanx, is an asynchronous array processing toolkit which
    transforms Python and NumPy operations into code which can be executed in
    parallel on HPC resources by mapping Python and NumPy functions and
    variables into a dependency tree executed by HPX, a general purpose,
    parallel, task-based runtime system written in C++. Phylanx additionally
    provides introspection and visualization capabilities for debugging and
    performance analysis. We have tested the foundations of our approach by
    comparing our implementation of widely used machine learning algorithms to
    accepted NumPy standards.


\end{abstract}


\section{Introduction}
\label{sec:intro}

The ever-increasing size of data sets in recent years have given the rise to
the term ``big data.'' The field of big data includes applications that utilize
data sets so large that traditional means of processing cannot handle
them~\cite{snijders2012big,doi:10.1108/LR-06-2015-0061}. The tools that
operate on such data sets are often termed as big data platforms. Some prominent
examples are Spark, Hadoop, Theano and
Tensorflow~\cite{HASHEM201598,6567202}.

One field which benefits form big data technology is Machine learning.
Machine learning techniques are used to extract useful data from these
large data sets~\cite{Landset2015,ALJARRAH201587}.
Theano~\cite{2016arXiv160502688short} and
Tensorflow~\cite{tensorflow2015-whitepaper} are two prominent frameworks that
support machine learning as well as deep learning~\cite{lecun2015deep} technology.
Both frameworks provide a Python interface, that has become the \textit{lingua franca}
for machine learning experts. This is due, in part, to the elegant math-like syntax
of Python that has been popular with domain scientists. Furthermore, the existence of
frameworks and libraries catering to machine learning in Python such as
NumPy, SciPy and Scikit-Learn have made Python the de facto standard
for machine learning.

While these solutions work well with mid-sized data sets, larger data sets still pose a
big challenge to the field.  Phylanx tackles this issue by
providing a framework that can execute arbitrary Python code in a distributed
setting using an asynchronous many-task runtime system. Phylanx is based on the open
source C++ library for parallelism and concurrency (HPX~\cite{Kaiser2014, Heller2017}).

This paper introduces the architecture of Phylanx and demonstrates how this solution enables
code expressed in Python to run in an HPC environment with minimal changes. While Phylanx
provides general distributed array functionalities that are applicable beyond the field
of machine learning, the examples in this paper focus on machine learning applications,
the main target of our research.

This paper makes the following contributions:
\begin{itemize}
    \item Describe the \textit{futurization} technique used to decouple the logical dependencies of the execution tree from its execution.
	\item Illustrate the software architecture of Phylanx.
	\item Demonstrate the tooling support which visualizes Phylanx's performance data to easily find bottlenecks and enhance
     performance.
	\item Present initial performance results of the method.
\end{itemize}

We will describe the background in Section~\ref{sec:background},
Phylanx's architecture in Section~\ref{sec:architecture}, study the performance of
several machine learning algorithms in Section~\ref{sec:experiments}, discuss related work
in Section~\ref{sec:related}, and present conclusions 
in Section~\ref{sec:conclusion}.

\section{Related Work}
\label{sec:related}
Because of the popularity of Python, there have been many efforts to
improve the performance of this language. Some specialized their solutions to
machine learning while others provide wider range of support for numerical
computations in general. NumPy~\cite{walt2011numpy} provides excellent
support for numerical computations on CPUs within a single node.
Theano~\cite{team2016theano} provides a syntax similar to NumPy, however, it
supports multiple architectures as the backend. Theano uses a symbolic
representation to enable a range of optimizations through its compiler.
PyTorch~\cite{paszke2017pytorch} makes heavy use of GPUs for high performance
execution of deep learning algorithms. Numba~\cite{lam2015numba} is a jit
compiler that speeds up Python code by using decorators. It makes use of LLVM
compiler to compile and optimize the decorated parts of the Python code. Numba
relies on other libraries, like Dask~\cite{dask} to support distributed
computation. Dask is a distributed parallel computation library implemented
purely in Python with support for both local and distributed executions of the
Python code. Dask works tightly with NumPy and Pandas~\cite{mckinneypandas} data
objects. The main limitation of Dask is that its scheduler has a per task
overhead in the range of few hundred microseconds, which limits its scaling
beyond a few thousand of cores. Google's
Tensorflow~\cite{tensorflow2015-whitepaper} is a symbolic math library with
support for parallel and distributed execution on many architectures and
provides many optimizations for operations widely used in machine learning.
Tensorflow is a library for dataflow programing which is a programming paradigm
not natively supported by Python and, therefore, not widely used.

\section{Technologies utilized to implement Phylanx}
\label{sec:background}
HPX~\cite{Kaiser2014, Heller2017} is an asynchronous many-task runtime system capable of running scientific applications both on a single process as well as in a distribued setting on thousands of nodes. HPX achieves a high degree of parallelism via lightweight tasks called HPX threads. These threads are scheduled on top of the Operating System threads via the
HPX scheduler, which implements an $M:N$ thread scheduling system. HPX threads can also be executed remotely via a form of active messages~\cite{von1992active} known as Parcels~\cite{wagle2018methodology,kaiser2009parallex}. We briefly introduce the technique of  \textit{futurization}, which is utilized within Phylanx. For more details we refer to~\cite{Heller2017}.\\

\begin{lstfloat}
\begin{lstlisting}[language=C++,frame=none,numbers=none,caption=Example for the concept of futurization within HPX. Example code was adapted from~\cite{kaiserblog15}.,label={code:future},escapechar=|]
//Definition of the function
int convert(std::string s){ return std::stoi(s); } |\label{code:f:def}|
//Asynchronous execution of the function
hpx::future<int> f = hpx::async(convert, "42"); |\label{code:f:async}|
//Accessing the result of the function
std::cout << f.get() << std::endl; |\label{code:f:print}|
\end{lstlisting}
\end{lstfloat}

The concept of futurization~\cite{kaiser2015higher} is illustrated in Listing~\ref{code:future}. The function in Line~\ref{code:f:def} is intended to be executed in parallel on one of the lightweight HPX threads. Line~\ref{code:f:async} shows the usage of the asynchronous return type \lstinline|hpx::future<T>|, the so-called \textit{Future}, of the asynchronous function call \lstinline|hpx::async|.  Note that \lstinline|hpx::async| returns the future immediately even though the computation within \lstinline|convert| may not have started yet. In Line~\ref{code:f:print}, the result of the future is accessed via its
member function \lstinline|.get()|. Listing~\ref{code:future} is just a simple usecase of futurization which does not
handle synchronization very efficiently. Consider the call to \lstinline|.get()|, if the Future has not become "ready"
\lstinline|.get()| will cause the current thread to suspend. Each suspension
will incur a context switch from the current
thread which adds overhead to the execution time. It is very important to avoid these
unnecessary suspensions for maximum efficiency.   \\

Fortunately, HPX provides barriers for the synchronization of dependencies. These include:
\lstinline|hpx::wait_any|, \lstinline|hpx::wait_any|, and  \lstinline|hpx::wait_all().then()|.
These barriers provide the user with a means to wait until a \lstinline|future| is ready before attempting to retrieve its value.
In HPX we have combined the \lstinline|hpx::wait_all().then()| facility and provided the user
with the \lstinline|hpx::dataflow| API~\cite{kaiser2015higher} demonstrated in Listing~\ref{code:dataflow}.

\begin{lstfloat}
\begin{lstlisting}[language=C++,frame=none,numbers=none,caption=Example for the concept of \lstinline|hpx::dataflow| for the transverse of a tree. Example code was adapted from~\cite{kaiserblog15}.,label={code:dataflow},escapechar=|]
template <typename Func>
future<int> traverse(node& n, Func && f)
{
    // traversal of left and right sub-tree
    future<int> left = |\label{code:d:left}|
        n.left ? traverse(*n.left, f)
        		: make_ready_future(0);
    future<int> right = |\label{code:d:right}|
        n.right ? traverse(*n.right, f)
        		: make_ready_future(0);

    // return overall result for current node
    return dataflow( |\label{code:d:data}|
        [&n, &f](future<int> l, future<int> r)
        		-> int
        {
            // calling .get() does not suspend
            return f(n) + l.get() + r.get(); |\label{code:d:get}|
        },
        std::move(left), std::move(right)
    );
}
\end{lstlisting}
\end{lstfloat}

Listing~\ref{code:dataflow} uses \lstinline|hpx::dataflow| to traverse a tree. In Line~\ref{code:d:left} and Line~\ref{code:d:right} the futures for the left and right traversal are returned. Note that these futures may have
not been computed yet when they are passed into the \lstinline|dataflow| on Line~\ref{code:d:data}.
The user could have used an \lstinline|hpx::async| here instead of \lstinline|hpx::dataflow|, but the Future passed
to the called function may have suspended the thread while waiting for its results in the \lstinline|.get()| function. The
\lstinline|hpx::dataflow| will not pass the Future arguments to the function until all of the Futures passed to the
\lstinline|hpx::dataflow| are "ready". This avoids the suspension of the child function call. In
Section~\ref{sec:architecture} \textit{futurization} and the facility \lstinline|hpx::dataflow| are heavily utilized to construct
the asynchronous architecture of Phylanx. \\

Finally, the last technology we used to guide the development of Phylanx is NumPy. NumPy~\cite{walt2011numpy} is a highly optimized numerical computation library for Python. NumPy is used in many scientific applications and supports highly performant, multidimensional array operations for scientific computing. Phylanx uses the library's API as the interface
to the user. In addition, Phylanx supports numerical computation on NumPy data objects through pybind11~\cite{pybind11} without the need for data copies.

\begin{figure}[hb]
	\begin{center}
		\begin{tikzpicture}
		[auto, scale=.78,
		every node/.style={rounded corners, scale =.78},
		line/.style ={draw, thick}
		]
		
		\node [minimum height=1cm,minimum width=6.2cm,draw=orange,text=orange,fill=orange!20, rotate=90] at (0,0.4){Visualization Tools};
		
		\node[minimum height=6.2cm,minimum width=4cm,draw=blue,fill=blue!30] at (3,0.4) {};
		\node [minimum height=1cm,minimum width=4cm,draw=none,text=blue ] at (3,3.8){Phylanx};
		
		\node[minimum height=1cm,minimum width=3.8cm,draw=teal,fill=teal!20,text=teal] at (3,2.9) {Frontend};
		\node[minimum height=1cm,minimum width=3.8cm,draw=teal,fill=teal!20,text=teal] at (3,1.9) {Optimizer};
		\node[minimum height=4cm,minimum width=3.8cm,draw=teal,fill=teal!20] at (3,-0.6) {};
		
		\node[minimum height=1cm,minimum width=3.8cm,draw=none,text=teal] at (3,0.9) {Backend};

		\node[minimum height=1cm,minimum width=3.2cm,draw=cyan,fill=cyan!20,text=cyan] at (3,0) {Compiler};
		\node[minimum height=2cm,minimum width=3.2cm,draw=cyan,fill=cyan!20] at (3,-1.5) (executor) {};
		\node[minimum height=1cm,minimum width=3.2cm,draw=none,text=cyan] at (3,-1) {Executor};
		\node[minimum height=1cm,minimum width=2.8cm,draw=red,fill=red!10,text=red] at (3,-1.9) {Perf. Counters};
		
		\node [minimum height=1cm,minimum width=6.2cm,draw=darkgray,text=darkgray,fill=darkgray!20, rotate=90] at (6,0.4)(pybind) {pybind11};
		
		\node[minimum height=1cm,minimum width=2cm,draw=darkgray,text=darkgray,fill=darkgray!20] at (8,2.5) {Python};
		\node[minimum height=1cm,minimum width=2cm,draw=darkgray,text=darkgray,fill=darkgray!20] at (8,0.5) {NumPy};
		\node[minimum height=1cm,minimum width=2cm,draw=darkgray,text=darkgray,fill=darkgray!20] at (8,-1.5) {Blaze};
		
		\node[minimum height=1cm,minimum width=8cm,draw=blue,fill=blue!20,text=blue] at (5,-3.5) {HPX};
		
		\draw [black, very thick] (3,-2.7) -- (3,-3);
		\draw [black, very thick] (8,-2) -- (8,-3);
		\draw [black, very thick] (6.5,-1.5) -- (7,-1.5);
		\draw [black, very thick] (6.5,0.5) -- (7,0.5);
		\draw [black, very thick] (6.5,2.5) -- (7,2.5);
		\draw [black, very thick] (4.6,-1.5) -- (5.5,-1.5);
		\draw [black, very thick] (4.6,0) -- (5.5,0);
	\end{tikzpicture}

	\end{center}
	\caption{Overview of the Phylanx toolkit and its interactions with external	libraries.}

	\label{fig:phylanxarch}
\end{figure}
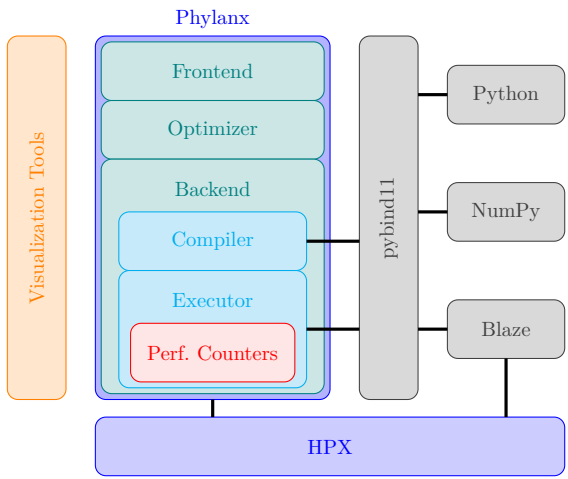

\begin{figure*}[!ht]
	\centering
	\includegraphics[width=1\linewidth]{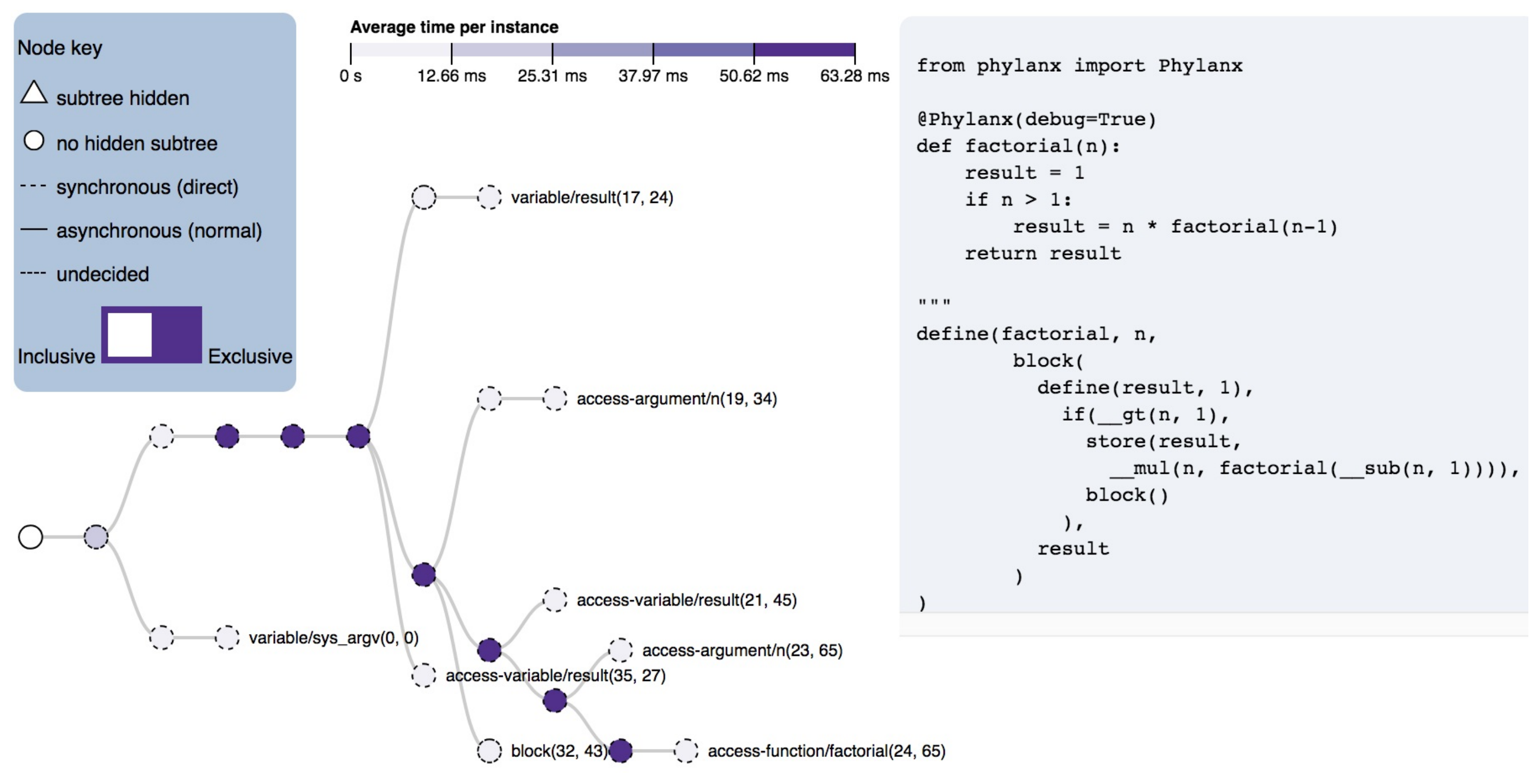}
	\par\caption{%
	Phylanx visualization tool provides a side-by-side view of the code and the
	corresponding expression tree along with performance information collected
	by builtin performance counters.}
	\label{fig:traveler_tree}
  \end{figure*}

\section{Phylanx}
\label{sec:architecture}
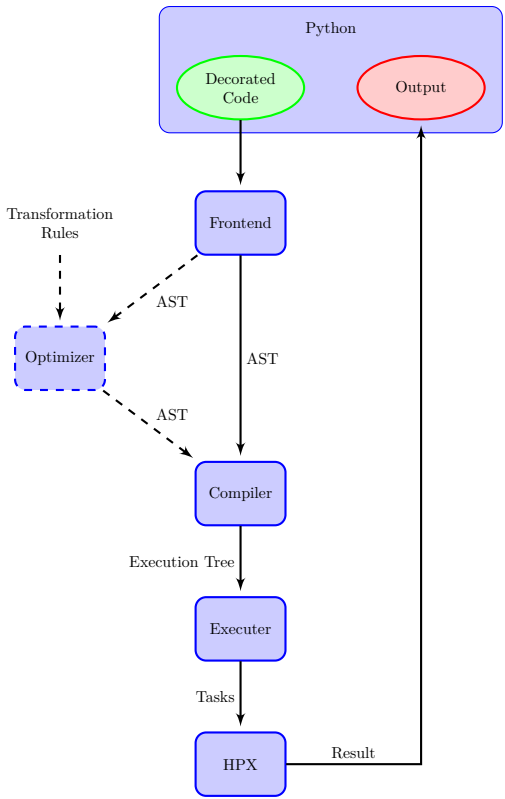
\begin{figure}[h]
	\begin{center}
	
	\begin{tikzpicture}
	[auto, scale=0.6,
	every node/.style={scale =0.6},
	ghost/.style ={rectangle, draw=none, text width=7em,
		align=center, rounded corners,
		minimum height=4em},
	input/.style ={	draw=green, thick, ellipse,fill=green!20,
		text width=5em,align=center,
		minimum height=4em},
	output/.style ={	draw=red, thick, ellipse,fill=red!20,
		text width=5em,align=center,
		minimum height=4em},
	block/.style 	={rectangle, draw=blue, thick, fill=blue!20,
		text width=5em,align=center, rounded corners,
		minimum height=4em},
	line/.style ={draw, thick, -latex',shorten >=2pt}
	]
	\node[minimum height=2.8cm,minimum width=7.6cm,draw=blue,fill=blue!20,rounded corners] at (2,15.4) {};
	\node [minimum height=1cm,minimum width=4cm,draw=none] at (2,16.3){Python};
	\node [input] (start) at (0,15) {Decorated Code};
	\node [output] (end) at (4,15)  {Output};
	\node [ghost] (middle) at(-4,12) {Transformation Rules};
	\node [block] (frontend) at (0,12) {Frontend};
	\node [block,dashed] (optimizer) at (-4,9) {Optimizer};
	\node [block] (compiler) at (0,6) {Compiler};
	\node [block] (executer) at (0,3)  {Executer};
	\node [block] (hpx) {HPX};
	
	\begin{scope}[every path/.style=line]
	\path 	(start) 	-- (frontend);
	\path 	(frontend) 	-- node [midway] {AST} (compiler);
	\path 		[dashed] (frontend) 	-- node [midway] {AST} (optimizer);
	\path 		[dashed] (optimizer) 	-- node [midway] {AST} (compiler);
	\path 	[dashed] (middle) -- (optimizer);
	\path 	(compiler) 	-- node [midway,left] {Execution Tree} (executer);
	\path 	(executer) 	-- node [midway,left] {Tasks} (hpx);
	\path 	(hpx) -| node [near start,above] {Result} (end);
	\end{scope}
	\end{tikzpicture}
	
	\end{center}
	\caption {Phylanx program flow. Phylanx frontend generates AST (PhySL) of the
	decorated Python code. The AST could be directly passed to the compiler to
	generate the execution tree or, optionally, fed to the optimizer first and then
	the compiler. Once the Kernel is invoked, Phylanx triggers the evaluation of the
	the execution tree on HPX. After finishing the evaluation, the result is
	returned in Python.}
	\label{fig:phylanxflow}
\end{figure}

In Python, the order of code blocks determines the execution order of a program
and implicit parallelism is only available within each block. Therefore,
asynchrony and parallelism across code blocks must be explicitly explored by the
programmer, a process which is tedious and error prone. In this section we discuss 
the implementation of our approach in Phylanx for automatic generation of task graphs and
the infrastructure used for running them on HPX for parallel, asynchronous,
distributed execution. We also discuss a suite of analysis and optimization
tools included in the Phylanx toolkit. Figure~\ref{fig:phylanxflow} provides an
overview of program flow in Phylanx.

\subsection{Frontend}
The Phylanx frontend provides two essential functionalities:

\begin{itemize}
	\item Transform the python code into a Phylanx internal representation called
	PhySL (Phylanx Specification Language).
	\item Copy-free handling of data objects between Python and Phylanx executor
	(in C++).
\end{itemize}

In addition, the frontend exposes two main functionalities of Phylanx that are
implemented in C++ and required for generation and evaluation of the execution
tree in Python. These functions are the \lstinline|compile| and
\lstinline|eval| methods.

\subsubsection{Code Transformation}
Performance benefits of many-task runtime systems, like HPX, are more prominent
when the compute load of the system exceeds the available resources. Therefore,
using these runtimes for sections of a program which are not computationally
intensive may result in little performance benefit. Moreover,
the overhead of code transformation and inherent extraneous work
imposed by runtime systems may even cause performance degradation.
Therefore, we have opted to limit our optimizations to performance critical
parts of the code which we call computational \textit{kernels}. The Phylanx frontend
provides a decorator (\lstinline|@Phylanx|) to trigger transformation of kernels into the
execution tree.

We have developed a custom internal representation of Python AST in order to
facilitate the analysis of static optimizations and streamline the generation of the
execution tree. The human-readable version of the AST, aka PhySL, is
automatically generated and compiled into the execution tree by the frontend.
More details on this compilation process can be found in~\ref{subsec:exe_tree}.
The benefit of
using PhySL as the intermediate representation is twofold: (1) it closely
reflects the nodes of the execution tree as each PhySL node represents a
function that will be run by an HPX task during evaluation, and (2) it can be
used for debugging and analyzing purposes for developers interested in custom
optimizations.

The compiled kernel is cached and can be be invoked directly in Python or in
other kernels.

\subsubsection{Data Handling}
\label{subsec:data}
Phylanx's data structures rely on the high-performance open-source C++ library
Blaze~\cite{iglberger2012expression,6266939}. Blaze already supports HPX as a
parallelization library backend and it perfectly maps its data to Python data
structures. Each Python list is mapped to a C++ \lstinline|vector| and 1-D and
2-D NumPy arrays are mapped to a Blaze vector and Blaze matrix respectively. To
avoid data copies between Python and C++, we take advantage of Python buffer
protocol through pybind11 library. Figure~\ref{fig:phylanxarch} shows how
Phylanx manages interactions with external libraries.

\subsection{Execution Tree}
\label{subsec:exe_tree}
After the transformation phase, the frontend passes the generated AST to the
Phylanx compiler to construct the execution tree where nodes are
\textit{primitives} and edges represent dependencies between parents and
children pairs.

Primitives are the cornerstones of the Phylanx toolkit and building blocks of the
Phylanx execution tree. Primitives are C++ objects which contain a single execute
function. This function is wrapped in a \lstinline|dataflow| and can
be as simple as a single instruction or as complex as a sophisticated algorithm.
We have implemented and optimized most Python constructs as well as many
NumPy methods as primitives. Futurization and asynchronous execution of tasks
are enabled through these constructs. One can consider primitives as lightweight
tasks that are mapped to HPX threads.
Each primitive accepts a list of futures as
its arguments and returns the result of its wrapped function as a future.
In this way, the primitive can accept both constant values known at compile time
as well as the results of previous primitives known only after being computed.

\subsection{Futurized Execution}
Upon the invocation of a kernel, Phylanx triggers the evaluation function of the
root node. This node represents the primitive corresponding to the
result of the kernel. In the evaluation function, the root node will call the
evaluation function of all of its children and those primitives will call the
evaluation functions of their children. This process will continue until the
the leaf nodes have been reached where the primitives evaluation functions
do not depend on other primitives to be resolved (e.g. a primitive which is a
constant, a primitive which reads from a file, etc.). It is important to note
that it does not matter where each primitive is placed in a distributed system
as HPX will resolve its location and properly call its eval function as well as
return the primitive's result to the caller.

As the leaf primitives are reached and their values, held in futures, are
returned to their parents the tree will unravel at the speed of the
critical path through the tree. The results from each primitive satisfy one of
the inputs of its parent node. After the root primitive finishes its execution,
the result of the entire tree is then ready to be consumed by the calling
function.

\subsection{Instrumentation}
Application performance analysis is a critical part of developing a parallel
application. Phylanx enables performance analysis by providing performance
counters to provide insight into its intrinsics. \textit{Time} performance
counters show the amount of time that is spent executing code in each subtree of
the execution tree, and \textit{count} performance counters show how many times
an execution tree node is executed. This data aides in identifying performance
hotspots and bottlenecks, which can either be directly used by the users or fed
into APEX\cite{huck2015autonomic} for adaptive load balancing. The data can also
be used by the visualization tools described in the next section.

\subsection{Visualization}

Embedding annotations and measurements for visualizations and performance
analysis within the runtime provides a
way to determine where performance bottlenecks are occurring and to gain insight
into the resource management within the machines. We show an example of
Phylanx's visualization capabilities in Figure ~\ref{fig:traveler_tree}. This
tree shows the execution tree from a test run of the factorial algorithm,
implemented in Python. In the tree, nodes are Phylanx primitives and edges show
parent/child relationships regarding how the child was called. The nodes are
colored purple for the \textit{inclusive time}, the total time spent executing
that primitive and its children. A switch in the toolbox in the upper left
corner allows for the user to switch from inclusive time to \textit{exclusive
time}, the time spent executing only that primitive. This allows for
identification of hotspots in the tree. Each primitive can be executed
asynchronously or synchronously in the parent thread. This distinction is shown
in dotted versus solid circles for nodes in the tree. The tree is interactive,
allowing users to drill down and focus by expanding or collapsing tree nodes and
hover for more details. The visualization is linked with a code view showing the
Python source code (the corresponding PhySL is shown as well). Hovering over a
node or line in one will highlight the corresponding line or node in the other.

\section{Experiments}
\label{sec:experiments}

\begin{figure*}[!ht]
  \centering
  \includegraphics[width=1\linewidth]{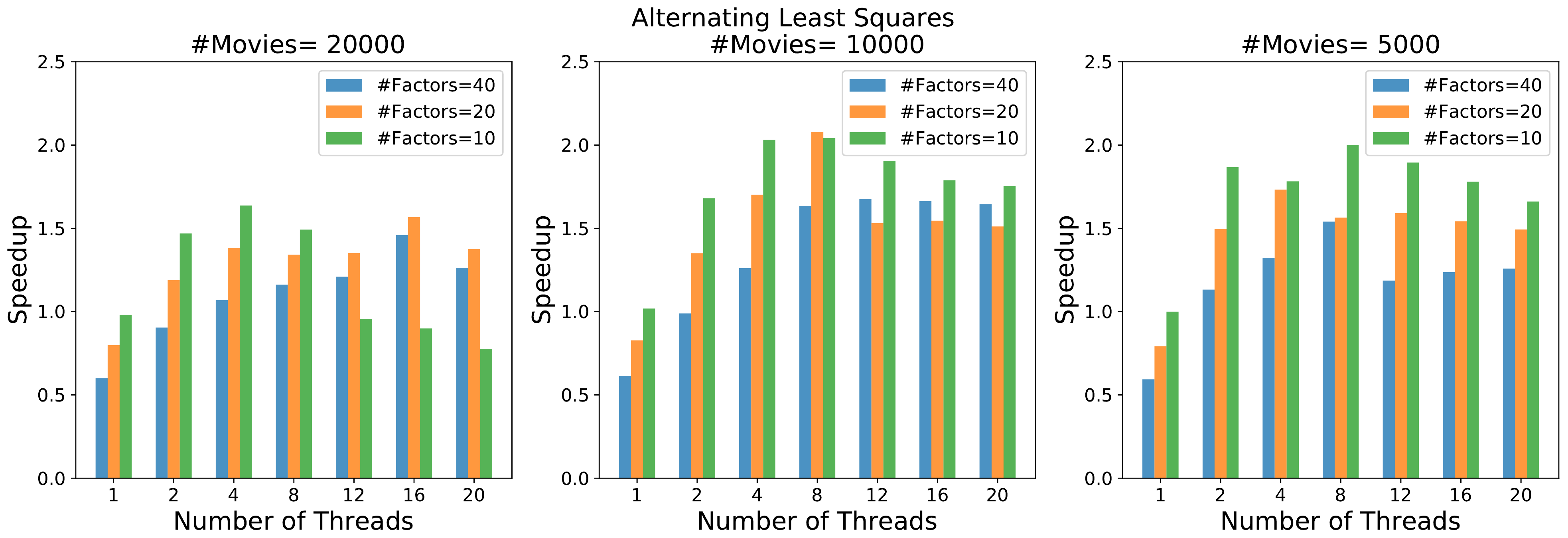}
  \par\caption{Speedup of the PhySL implementation of the ALS algorithm over the Python implementation. Number of threads represents the number of OS threads used by both Phylanx and Python. }
  \label{fig:als_bar}
\end{figure*}

This section details the performance comparison of PhySL to a corresponding
Python implementation utilizing multiple cores on top of NumPy using OpenBlas for BLAS/LAPACK routines. We used reference implementations of
Alternating Least Squares~\cite{hu2008collaborative} and Binary
Logistic Regression~\cite{bishop1} algorithms to analyze the performance of
equivalent code written in PhySL. The Logistic Regression coupled with Alternating Least Squares provides a wide variety of computationally intensive operations which makes them useful for experimentation and are also used as benchmarks for the Intel MKL Library~\cite{intelalslra}.
\subsection{Experimental Testbed}
We ran our experiments on LSU's Rostam cluster. These experiments were performed
on a node consisting of Intel(R) Xeon(R) CPU E5-2660 v3 clocked at 2.6GHZ,
with 10 cores (20 threads), and 128 GB DDR4 Memory. All Experiments
were performed on HPX v1.2 commit 9182ac6182, Phylanx v0.1 commit 116c46a8
Python v3.5.1, NumPy v1.15.0 , OpenBlas v0.3.2 and Blaze v3.3.

\subsection{LRA}
We implemented the Binary Logistic Regression algorithm in Python
and used the Phylanx decorator to generate the corresponding PhySL code. In
order to test the performance of the two implementations of the Logistic
regression algorithm, we created a custom binary classification
dataset with 10,000 features and 10,000 observations.

\begin{figure}[tbp]
  \centering
  \includegraphics[width=1\linewidth]{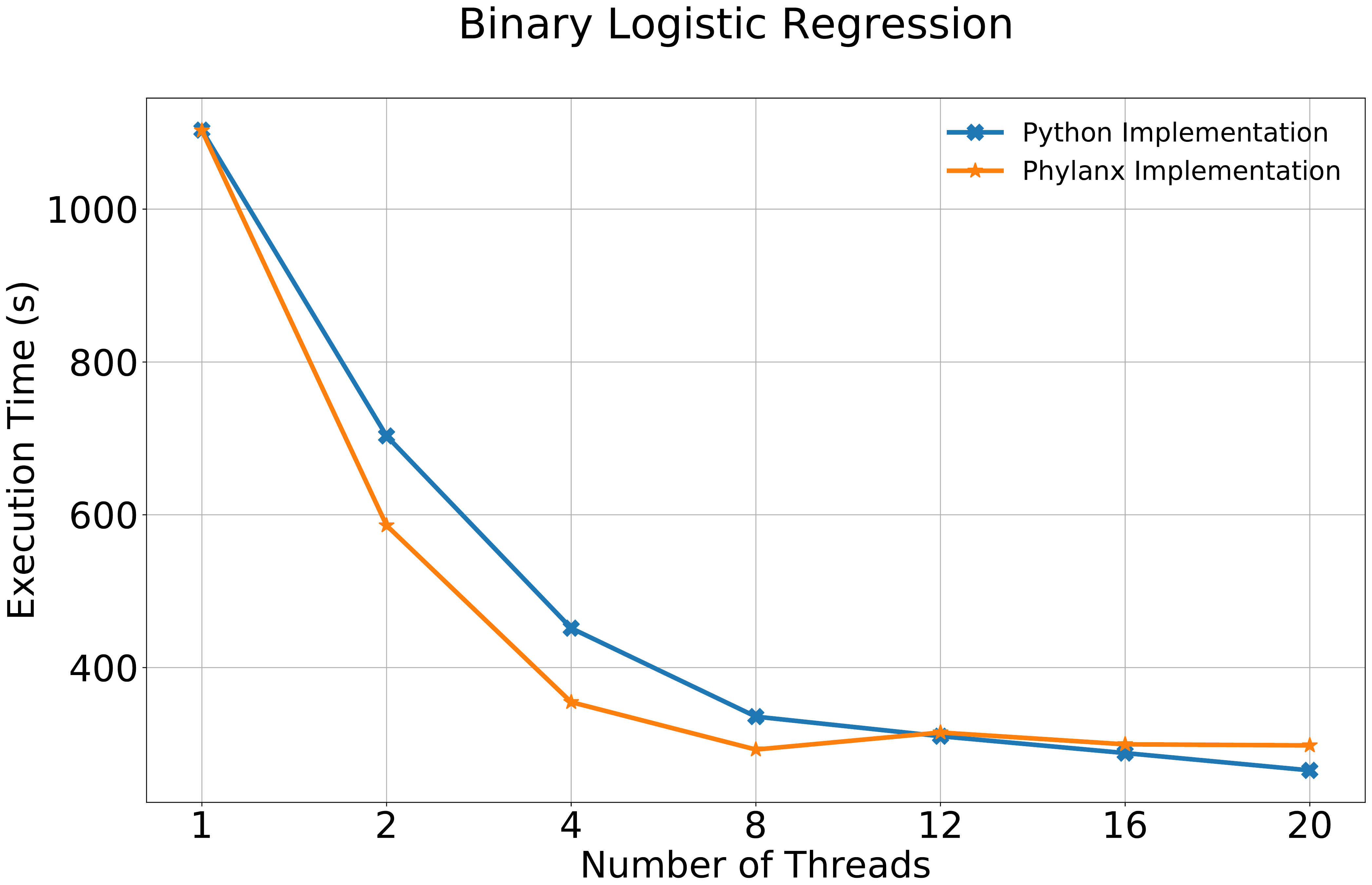}
  \par\caption{Comparing execution time of the reference implementation of
  the Logistic Regression algorithm in Python with the corresponding
  PhySL code. Each datapoint represents
  the average execution time over ten runs.}
  \label{fig:purephysl_numpy_phylanx_lra}
\end{figure}

Figure~\ref{fig:purephysl_numpy_phylanx_lra} shows the performance of the Python
and PhySL codes in terms of the execution time. Our experiments show that on a
single thread both PhySL and Python perform on par with each other. However,
PhySL scales faster up to eight cores and plateaus afterwards while Python
scales at a lower rate but up to 20 cores.

\subsection{Alternating Least Squares}
Alternating Least Squares is a method used in collaborative filtering based on
matrix factorization~\cite{hu2008collaborative}. Collaborative filtering as a
recommender system is utilized to predict a user's interest in a set of items
based on other users interaction with those items, and also the user's
interactions with other items. In order to test the implementation of the Alternating Least Squares algorithm in Phylanx, we implemented the algorithm in Python using NumPy and generated the corresponding PhySL implementation using the Phylanx decorator .
The two implementations of the algorithms were tested on
MovieLens-20M dataset~\cite{movielens}, which is a
collection of 20 million ratings gathered by 138,000 users over 27,000 movies.

\begin{figure}[tpb]
  \centering
  \includegraphics[width=1\linewidth]{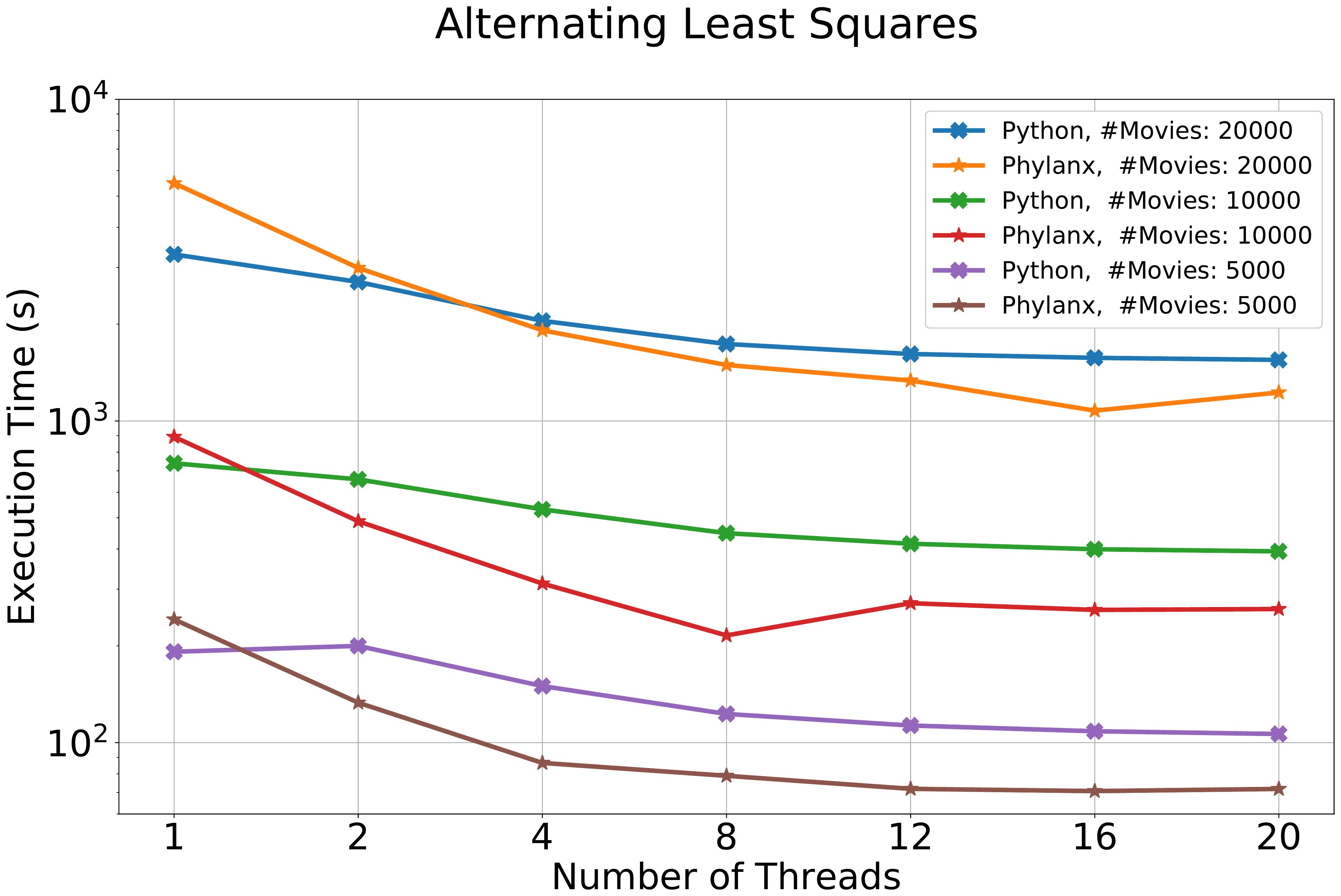}
  \par\caption{Comparing execution time of the reference implementation of
  the Alternating Least Squares algorithm in Python and the
  corresponding PhySL code. Each datapoint
  represents the average execution time over ten runs.}
  \label{fig:purephysl_numpy_phylanx_als}
\end{figure}

Figure~\ref{fig:purephysl_numpy_phylanx_als} shows the execution times of the PhySL and Python versions of Alternating least squares. Both versions were run with number of factors set to 40 while the number of movies were set to 5,000, 10,000 and 20,000.
The PhySL version of the ALS algorithm starts to outperform the
Numpy/Python version as the number of threads increases. The fastest time for
the Phylanx implementation is seen using 16 threads and the number of movies set to
20,000. When the number of movies were set to 10,000, the fastest time was seen using
12 threads. There is a noticeable
difference in execution time between the Python and the PhySL implementation
when the number of threads is set to one. In this configuration the Phylanx version is
much slower than the Python implementation. Such behavior is not seen with the
Logistic Regression example. This behavior is currently under investigation.

Figure~\ref{fig:als_bar} shows the speedup of the Phylanx implementation of
the Alternating least square using the
Python implementation's performance as the baseline. Both implementations
were run with the number of factors set
to 10, 20 and 40. while the number of movies were set to 5,000, 10,000, and 20,000.
It is seen that the Phylanx implementation outperforms the Python
implementation as the number of threads are increased on wide variety of problem sizes.
\section{Conclusion}
\label{sec:conclusion}
Despite the solutions provided by current machine learning frameworks,
better methodologies are needed to process the large amounts of data
consumed by cutting edge machine learning applications in a timely manner.
These new tools need to be accessible to the domain scientists who currently
use them as well as efficient with the computational resources provided to them.

In this paper, we have introduced a novel approach for transformation and
execution of high-productivity languages on top of the highly performant,
low-level HPX runtime system. We have implemented our approach along with a
suit of performance and visualization tools in the Phylanx array processing
toolkit. Phylanx enables automatic generation of asynchronous task graphs from
regular Python code and facilitates finer grain configurability. Our early experiments
on representative applications and datasets demonstrate the performance benefits of
our methods on a single node. However, we expect that the real benefits
of our approach will manifest in a distributed runtime environment.
Here the asynchronous execution and locality abstractions provided by
HPX stand to benefit users immensely by drastically decreasing execution times 
with little effort from the user.

\section{Future work}
\label{sec:futurework}
As the Phylanx technology matures we intend focus on two major goals:
first, to improve the performance of single node runs and second to 
extend the framework to automatically run user supplied codes in
distributed settings. Improving single node runs will entail implementing more 
basic algorithms utilized by the machine learning community and 
using that experience to improve the underlying toolkit. We anticipate
that we will be able to uncover performance bugs as well as opportunities
to improve the performance of our toolkit from this experience. One
such opportunity is the support for hardware accelerators such as GPUs.

Phylanx plans to enable users to execute their existing Python code on
clusters. This provides them with the ability to handle large data sets and
achieve better scaling. While Phylanx is built with distributed runs in mind
(execution trees can span across several nodes and evaluation is done using
features from HPX), distributed runs will require extensions to the current
toolkit to 
determine optimal data layout and data tiling given the algorithms
provided by the user. We also intend to look at using code transformations
to replace slower user-written algorithms with more efficient ones. 
While these goals present a research challenge, we believe
that the support provided by the HPX runtime system will substantially reduce 
the barriers to providing distributed execution capabilities.

\section{Acknowledgments}
This work was funded by the NSF Phylanx project award \#1737785. Any opinions,
findings, and conclusions or recommendations expressed in this material are
those of the authors and do not necessarily reflect the views of the National
Science Foundation.
%



\bibliographystyle{IEEEtran}
\bibliography{references}
\clearpage

\end{document}